\documentstyle [12pt,twoside, epsf]{article}
\let\\=\cr

\def\Chi{\hbox{\raise0.5ex\hbox{$\chi$}}}

\def\picill#1by#2(#3){\epsffile{#3}}

\begin{document}
\pagestyle{myheadings}

\title{\bf Biologic}

\author{Louis H. Kauffman}  


\date{}

\maketitle


\begin{abstract} 
 In this paper we explore the boundary between biology and the study of formal systems (logic).
\end{abstract}

\section{Introduction} 

This paper concentrates on relationships of formal systems with biology. In particular, this is a study of different forms
and formalisms for replication.  
\smallbreak  

\noindent
In living systems there is an essential circularity that is
the living structure. Living systems produce themselves from themselves and the materials and energy of the environment.
There is a strong contrast in how we avoid circularity in mathematics and how nature revels in biological circularity.
One meeting point of biology and 
mathematics is knot theory and topology. This is no accident, since topology is indeed a controlled study of cycles and circularities
in primarily geometrical systems.  
\bigbreak

In this paper we will discuss DNA replication, logic and biology, the relationship of symbol and object, the emergence of form. It is in 
the replication of DNA that the polarity (yes/no, on/off, true/false) of logic and the continuity of topology meet. Here 
polarities are literally fleshed out into the forms of life. 
\bigbreak

We shall pay attention 
to the different contexts for the logical, from the mathematical to the biological to the quantum logical. In each case
there is a shift in the role of certain key 
concepts. In particular, we follow the notion of copying through these contexts and with it gain new insight into the role of replication
in biology, in formal systems and in the quantum level (where it does not exist!). 
\bigbreak

In the end we arrive at a summary formalism, a chapter in {\em boundary mathematics} (mathematics using directly
the concept and notation of containers
and delimiters of forms - compare \cite{WB} and \cite{GSB}) where there are not only containers $<>$, but also
extainers \, $><$ -- entities open to interaction and distinguishing 
the space that they are not. In this formalism we find a key for the articulation of diverse relationships. The 
{\em boundary algebra of containers and extainers} is to biologic what boolean algebra is to classical logic. Let $C=<>$ and $E=><$ then 
$EE = ><>< = >C<$ and $CC=<><>=<E>$ Thus an extainer produces a container when it interacts with itself, and a container produces an extainer
when it interacts with itself.
\bigbreak

The formalism of containers and extainers is a chapter in the foundations of a symbolic language for shape and interaction. 
With it, we can express the {\em form} of DNA replication succinctly as follows: Let the DNA itself be represented as a container

$$\mbox{\rm DNA} = <>.$$
 
\noindent We regard the two brackets of the container as representatives for the two matched DNA strands. We let the extainer $E=><$ represent the 
cellular environment with its supply of available base pairs (here symbolized by the individual left and right brackets).  Then when the DNA
strands separate, they encounter the matching bases from the environment and become two DNA's.

$$ \mbox{\rm DNA} = \, <> \, \longrightarrow \, < E > \, \longrightarrow \, < > < > \, = \, \mbox{\rm DNA} \,\, \mbox{\rm DNA}.$$ 

\noindent
Life itself is about systems that search and learn and become. Perhaps a little symbol like $E =><$ with the property that 
$EE = ><><$ produces containers $<>$ and retains its own integrity in conjunction with the autonomy of $<>$ (the DNA)
could be a step toward bringing formalism to life.
\bigbreak

\noindent {\bf Acknowledgment.} The author thanks Sofia Lambropoulou for many useful conversations
in the course of preparing this paper. The author also thanks Sam Lomonaco, John Hearst, Yuri Magarshak, James Flagg and William Bricken
for conversations related to the content of the present paper.
\smallbreak

\noindent Most of this effort was sponsored by the Defense
Advanced Research Projects Agency (DARPA) and Air Force Research Laboratory, Air
Force Materiel Command, USAF, under agreement F30602-01-2-05022. 
The U.S. Government is authorized to reproduce and distribute reprints
for Government purposes notwithstanding any copyright annotations thereon. The
views and conclusions contained herein are those of the author and should not be
interpreted as necessarily representing the official policies or endorsements,
either expressed or implied, of the Defense Advanced Research Projects Agency,
the Air Force Research Laboratory, or the U.S. Government. (Copyright 2002.)  
\bigbreak

\section{Replication of DNA}
We start this essay with the question: During the replication of DNA, how do the daughter
DNA duplexes avoid entanglement? In the words of John Hearst \cite{KH}, we are in search of the mechanism for
the ``emaculate segregation".  This question is inevitably involved with the topology of the DNA, for
the strands of the DNA are interwound with one full turn for every ten base pairs. With the strands
so interlinked it would seem impossible for the daughter strands to separate from their parents.
\bigbreak

A key to this problem certainly lies in the existence of the topoisomerase enzymes that can change the
linking number between the DNA strands and also can change the linking number between two DNA duplexes.
It is however, a difficult matter at best to find in a tangled skein of rope the just right crossing 
changes that will unknot or unlink it.  The topoisomerase enzymes do just this, changing crossings by 
grabbing a strand, breaking it and then rejoining it after the other strand has slipped through
the break. Random strand switching is an unlikely mechanism, and one is led to posit some intrinsic
geometry that can promote the process. In \cite{KH} there is made a specific suggestion about this intrinsic 
geometry. It is suggested that in vivo the DNA polymerase enzyme that promotes replication (by creating 
loops of single stranded DNA by opening the double stranded DNA) has sufficient rigidity not to allow the new 
loops to swivel and become entangled. In other words, it is posited that the replication 
loops remain simple in their topology so that the topoisomerase can act to promote the formation of 
the replication loops, and these loops once formed do not hinder the separation of the newly born 
duplexes. The model has been to some degree confirmed \cite{LD}.  The situation would now appear to be that
in the first stages of the formation of the replication loops Topo\,I acts favorably to allow their 
formation and amalgamation. Then Topo\,II has a smaller job of finishing the separation of the newly 
formed duplexes. In Figure 1 we illustrate the schema of this process. In this Figure we indicate the action of the Topo\,I
by showing a strand being switched in between two replication loops. The action of Topo\, II is only stated but not shown.
In that action, newly created but entangled DNA strands would be disentangled. Our hypothesis is that this second action is
essentially minimized by the rigidity of the ends of the replication loops in vivo.
\bigbreak

$$ \picill5inby5.5in(Figure1)  $$

\begin{center}
{\bf Figure 1  - DNA Replication }
\end{center}

In the course of this research, we started thinking about the diagrammatic logic of DNA replication
and more generally about the relationship between DNA replication, logic and basic issues in the 
foundations of mathematics and modeling. The purpose of this paper is to explain some of these 
issues, raise questions and place these questions in the most general context that we can muster at
this time. The purpose of this paper is therefore foundational.  It will not in its present form
affect issues in practical biology, but we hope that it will enable us and the reader to ask fruitful
questions and perhaps bring the art of modeling in mathematics and biology forward.
\bigbreak

To this end we have called the subject matter of this paper ``biologic" with the intent that this 
might suggest a quest for the logic of biological systems or a quest for a ``biological logic" or 
even the question of the relationship between what we call ``logic" and our own biology. We 
have been trained to think of physics as the foundation of biology, but it is possible to 
realize that indeed biology can also be regarded as a foundation for thought, language, mathematics 
and even physics. In order to bring this statement over to physics one has to learn to admit that
physical measurements are performed by biological organisms either directly or indirectly and that
it is through our biological structure that we come to know the world. This foundational view will be 
elaborated as we proceed in this paper.
\bigbreak

\section{Logic, Copies and DNA Replication}
In logic it is implicit at the syntactical level that copies of signs are 
freely available. In abstract logic there is no issue about materials available for the production of copies of
a sign, nor is there necessarily a formalization of how a sign is to be copied. In the practical realm there
are limitations to resources. A mathematician may need to replenish his supply of paper. A computer
has a limitation on its memory store. In biology, there are no signs, but there are entities that 
we take as signs in our description of the workings of the biological information process. In this 
category the bases that line the backbone of the DNA are signs whose significance lies in their 
relative placement in the DNA. The DNA itself could be viewed as a text that one would like to copy.
If this were a simple formal system it would be taken for granted that copies of any given text can
be made. Therefore it is worthwhile making a comparison of the methods of copying or reproduction that
occur in logic and in biology.
\bigbreak

In logic there is a level beyond the simple copying of symbols that contains a non-trivial description
of self-replication.  The schema is as follows: There is a universal building machine $B$ that 
can accept a text or description $x$ (the program) and build what the text describes. We let lowercase
$x$ denote the description and uppercase $X$ denote that which is described.  Thus $B$ with $x$ will build
$X.$  In fact, for bookkeeping purposes we also produce an extra copy of the text $x.$ This is 
appended to the production $X$ as $X,x.$ Thus $B$, when supplied with a description $x,$ produces that which $x$
describes, with a copy of its description attached. Schematically we have the process shown below.
\bigbreak

$$B,x \longrightarrow B,x ; X,x$$

\noindent
Self-replication is an immediate consequence of this concept of a universal building machine.
Let $b$ denote the text or program for the universal building machine.  Apply $B$ to
its own description.

$$B,b \longrightarrow B,b ; B,b$$

\noindent
The universal building machine reproduces itself. Each copy is a universal 
building machine with its own description appended. Each copy will proceed to reproduce itself
in an unending tree of duplications. In practice this duplication will continue until all available
resources are used up, or until someone removes the programs or energy sources from the 
proliferating machines.
\bigbreak

It is not necessary to go all the way to a universal building machine to establish 
replication in a formal system or a cellular automaton (See the epilogue to this paper for examples.).
On the other hand, all these logical devices for replication
are based on the hardware/software or Object/Symbol distinction. It is worth looking at the abstract 
form of DNA replication.
\bigbreak

DNA consists in two strands of base-pairs wound helically around a phosphate backbone.  It is 
customary to call one of these strands the ``Watson" strand and the other the ``Crick" strand.
Abstractly we can write  $$DNA = <W|C>$$ \noindent to symbolize the binding of the two strands into the single
DNA duplex. Replication occurs via the separation of the two strands via polymerase enzyme.
This separation occurs locally and propagates. Local sectors of separation can amalgamate into
larger pieces of separation as well. Once the strands are separated, the environment of the cell 
can provide each with complementary bases to form the base pairs of new duplex DNA's. Each strand,
separated in vivo, finds its complement being built naturally in the environment. This picture ignores
the well-known topological difficulties present to the actual separation of the daughter strands.
\bigbreak

The base pairs are $AT$ (Adenine and Thymine) and $GC$ (Guanine and Cytosine). Thus if
 
$$<W| = <...TTAGAATAGGTACGCG... |$$ \noindent Then $$|C> = |...AATCTTATCCATGCGC...>.$$ 

\noindent Symbolically we can oversimplify the whole process as 

$$<W| + E \longrightarrow <W|C> = DNA$$

$$E + |C> \longrightarrow <W|C> = DNA$$

$$<W|C> \longrightarrow <W| + E + |C> = <W|C> <W|C>$$

\noindent Either half of the DNA can, with the help of the environment, become a full DNA.
We can let $E \, \longrightarrow \, |C><W|$ be a symbol for the process by which the environment supplies the 
complementary base pairs $AG$, $TC$ to the Watson and Crick strands. In this oversimplification
we have cartooned the environment as though it contained an already-waiting strand $|C>$ to 
pair with $<W|$ and an already-waiting strand $<W|$ to pair with $|C>.$ 
\smallbreak 

\noindent {\em In fact it is the opened
strands themselves that command the appearance of their mates. They conjure up their mates from 
the chemical soup of the environment.} 
\smallbreak 

\noindent The environment $E$ is an identity element in this algebra of 
cellular interaction. That is, $E$ is always in the background and can be allowed to appear spontaneously
in the cleft between Watson and Crick:

$$<W|C> \longrightarrow  <W| |C> \longrightarrow <W| E |C>$$ 

$$\longrightarrow <W| |C><W| |C> \longrightarrow <W|C><W|C>$$

\noindent This is the formalism of DNA replication.  
\bigbreak

Compare this method of replication with the movements of the universal building machine supplied 
with its own blueprint. Here Watson and Crick ( $<W|$ and $|C>$ ) are each both the machine {\em and} 
the blueprint for the DNA. They are complementary blueprints, each containing the information to
reconstitute the whole molecule. They are each machines in the context of the cellular 
environment, enabling the production of the DNA.  This coincidence of machine and blueprint, hardware
and software is an important difference between 
classical logical systems and the logical forms that arise in biology.
\bigbreak

\section{Lambda Algebra - Replication Revisited}
One can look at formal systems involving self-replication that do not make a 
distinction between Symbol and Object. In the case of formal systems this means that one is 
working entirely on the symbolic side, quite a different matter from the biology where there is no
intrinsic symbolism, only our external descriptions of processes in such terms. An example 
at the symbolic level is provided by the lambda calculus of Church and Curry \cite{BD} where functions
are allowed to take themselves as arguments. This is accomplished by the following axiom. 
\bigbreak

\noindent {\bf Axiom for a Lambda Algebra}:  Let $A$ be an algebraic system with one binary operation denoted 
$ab$ for elements $a$ and $b$ of $A.$ Let $F(x)$ be an algebraic expression over $A$ with one variable $x.$
Then there exists an element $a$ of $A$ such that $F(x) = ax$ for all $x$ in $A.$ 
\bigbreak

An algebra (not associative) that satisfies this axiom is a representation of the
lambda calculus of Church and Curry.  Let $b$ be an element of $A$ and define $F(x) = b(xx).$  Then 
by the axiom we have $a$ in $A$ such that  $ax = b(xx)$ for any $x$ in $A.$ In particular (and this is where
the ``function" becomes its own argument)  $$aa = b(aa).$$ Thus we have shown that for any $b$ in $A$, there exists
an element $x$ in $A$ such that $x = bx.$ Every element of $A$ has a ``fixed point."

\noindent
This conclusion has two effects. It provides a fixed point for the function $G(x) = bx$ and it creates the
beginning of a recursion in the form 

$$aa = b(aa) = b(b(aa)) = b(b(b(aa))) = ...$$

\noindent The way we arrived at the fixed point $aa$ was formally the same as the
mechanism of the universal building machine.  Consider that machine:
\bigbreak

$$B,x \longrightarrow X,x$$   

\noindent We have left out the repetition of the machine itself. You could look at this as a machine that 
uses itself up in the process of building $X.$ Applying $B$ to its own description $b$ we have
the self-replication

$$B,b \longrightarrow B,b.$$

\noindent The repetition of $x$ in the form $X,x$ on the right hand side of this definition of the 
builder property is comparable with  

$$ax = b(xx)$$

\noindent with its crucial repetition as well. In the fixed point theorem, the arrow is replaced by an 
equals sign! Repetition is the core of self-replication in classical logic.
{\em This use of repetition assumes the possibility of a copy at the syntactic level, in order to produce
a copy at the symbolic level.} There is, in this pivot on syntax, a deep 
relationship with other fundamental issues in logic. In particular this same form of repetition
is in back of the Cantor diagonal argument showing that the set of subsets of a set 
has greater cardinality than the original set, and it is in back of the G$\ddot{o}$del Theorem on the 
incompleteness of sufficiently rich formal systems. The pattern is also in back of the production of 
paradoxes such as the Russell paradox of the set of all sets that are not members of themselves.
\bigbreak

There is not space here to go into all these relationships, but the Russell paradox will give 
a hint of the structure.  Let ``ab" be interpreted as ``b is a member of a". Then $RX = \neg (XX)$ can be 
taken as the definition of a set $R$ such that $X$ is a member of $R$ exactly when it is {\em not} the case that 
$X$ is a member of $X.$ Note the repetition of $X$ in the definition $RX = \neg (XX).$ Substituting $R$ for $X$
we obtain  $RR = \neg (RR)$, which says that $R$ is a member of $R$ exactly when it is not the case that 
$R$ is a member of $R.$ This is the Russell paradox. From the point of view of the lambda calculus, we have found a 
fixed point for negation.
\bigbreak

Where is the repetition in the DNA self-replication?  The repetition and the replication are no longer separated. The repetition 
occurs not syntactically, but directly at the point of replication. Note the device of 
pairing or mirror imaging.  $A$ calls up the appearance of $T$ and $G$ calls up the appearance of $C.$
$<W|$ calls up the appearance of $|C>$ and $|C>$ calls up the appearance of $<W|.$  Each object $O$ calls up 
the appearance of its {\em dual or paired object} $O^*$.  $O$ calls up $O^*$ and $O^*$ calls up $O.$  The object that 
replicates is implicitly a repetition in the form of a pairing of object and dual object.
\smallbreak

\noindent $OO^*$ replicates via

$$O \longrightarrow OO^*$$

$$O^* \longrightarrow OO^*$$

\noindent whence

$$OO^* \longrightarrow O \,\,\, O^* \longrightarrow OO^* \,\,\, OO^*.$$

\noindent The repetition is inherent in the replicand in the sense that the dual of a form is a repetition of 
that form.
\bigbreak

\section{Quantum Mechanics}
We now consider the quantum level. Here copying is not
possible. We shall detail this in a subsection. For a quantum process to copy a state, one needs a
unitary transformation to perform the job. One can show, as we explain in the last subsection of this section, that this 
cannot be done. There are indirect ways that seem to make a copy, involving
a classical communication channel coupled with quantum operators (so called quantum teleportation \cite{Lom}).
The production of such a quantum state constitutes a reproduction of the original
state, but in these cases the original state is lost, so teleportation looks more like transportation than copying.  
With this in mind it is fascinating to contemplate that DNA and other molecular configurations
are actually modeled in principle as certain complex quantum states.  At this stage we meet the
boundary between classical and quantum mechanics where conventional wisdom finds it is most useful to regard the main level of  
molecular biology as classical.
\bigbreak

We shall quickly
indicate the basic principles of quantum mechanics.  The quantum information context 
encapsulates a concise model of quantum theory:
\bigbreak

{\em The initial state of a quantum process is a vector $|v>$ in a complex vector space $H.$
Observation returns basis elements $\beta$ of $H$ with probability 

$$|<\beta \,|v>|^{2}/<v \,|v>$$

\noindent where $<v \,|w> = v^{*}w$ with $v^{*}$ the conjugate transpose of $v.$
A physical process occurs in steps $|v> \longrightarrow U\,|v> = |Uv>$ where $U$ is a unitary linear transformation.
\bigbreak

Note that since $<Uv \,|Uw> = <v \,|w>$ when $U$ is unitary, it follows that probability is preserved in the 
course of a quantum process.  }
\bigbreak

One of the details for any specific quantum problem is the nature of the unitary 
evolution.  This is specified by knowing appropriate information about the classical physics that 
supports the phenomena. This information is used to choose an appropriate Hamiltonian through which the 
unitary operator is constructed via a correspondence principle that replaces classical variables with appropriate quantum
operators. (In the path integral approach one needs a Langrangian to construct the action on which the path
integral is based.) One needs to know certain aspects of classical physics to 
solve any given quantum problem.  The classical world is known through our biology. In this sense 
biology is the foundation for physics.
\bigbreak

A key concept in the quantum information viewpoint is the notion of the superposition of states.
If a quantum system has two  distinct states $|v>$ and $|w>,$ then it has infinitely many states of the form
$a|v> + b|w>$ where $a$ and $b$ are complex numbers taken up to a common multiple. States are ``really" 
in the projective space associated with $H.$ There is only one superposition of a single state $|v>$ with 
itself. 
\bigbreak

Dirac \cite{D} introduced the ``bra -(c)-ket" notation $<A\,|B>= A^{*}B$ for the inner product of complex vectors $A,B \in H$.
He also separated the parts of the bracket into the {\em bra} $<A\,|$ and the {\em ket} $|B>.$ Thus

$$<A\,|B> = <A\,|\,\,|B>$$

\noindent In this interpretation,
the ket $|B>$ is identified with the vector $B \in H$, while the bra $<A\,|$ is regarded as the element dual to $A$ in the 
dual space $H^*$. The dual element to $A$ corresponds to the conjugate transpose $A^{*}$ of the vector $A$, and the inner product is 
expressed in conventional language by the matrix product $A^{*}B$ (which is a scalar since $B$ is a column vector). Having separated the bra and the ket, Dirac can write the
``ket-bra"  $|A><B\,| = AB^{*}.$ In conventional notation, the ket-bra is a matrix, not a scalar, and we have the following formula for the 
square of $P = |A><B\,|:$

$$P^{2} =  |A><B\,| |A><B\,| = A(B^{*}A)B^{*} = (B^{*}A)AB^{*} = <B\,|A>P.$$

\noindent Written entirely in Dirac notation we have
    
$$P^{2} =  |A><B\,| |A><B\,| =  |A><B\,|A><B\,|$$ 

$$= <B\,|A>\,|A\,><B| = <B\,|A>P.$$

\noindent The standard example is a ket-bra $P = |A\,><A|$ where $<A\,|A>=1$ so that $P^2 = P.$  Then $P$ is a projection matrix, 
projecting to the subspace of $H$ that is spanned by the vector $|A>$. In fact, for any vector $|B>$ we have 

$$P|B> = |A><A\,|\,|B> =  |A><A\,|B> = <A\,|B>|A>.$$

\noindent If $\{|C_{1}>, |C_{2}>, \cdots |C_{n}> \}$ is an orthonormal basis for $H$, and $P_{i} = |C_{i} \,><C_{i}|,$
then for any vector $|A>$ we have

$$|A> = <C_{1}\,|A>|C_{1}> + \cdots + <C_{n}\,|A>|C_{n}>.$$

\noindent Hence 

$$<B\,|A> = <C_{1}\,|A><B\,|C_{1}> + \cdots + <C_{n}\,|A><B\,|C_{n}>$$

$$ = <B\,|C_{1}><C_{1}\,|A> + \cdots + <B\,|C_{n}><C_{n}\,|A>$$

$$ = <B\,|\,\,\,[|C_{1}><C_{1}\,| + \cdots + |C_{n}><C_{n}\,|]\, \,\,|A>$$

$$ = <B\,|\,1 \,\,|A>.$$

\noindent We have written this sequence of equalities from $<B\,|A>$ to $<B\,|1\,|A>$ to emphasize the role of the identity

$$\Sigma_{k=1}^{n} P_{k} = \Sigma_{k=1}^{n} |C_{k}><C_{k}\,| = 1$$

\noindent so that one can write

$$<B\,|A> = <B\,|\,1 \,|A> = <B\,| \Sigma_{k=1}^{n} |C_{k}><C_{k}\,| |A> = \Sigma_{k=1}^{n} <B\,|C_{k}><C_{k}\,|A>.$$

In the quantum context
one may wish to consider the probability of starting in state $|A>$ and ending in state $|B>.$ The 
square of the probability for this event is equal to $|<B\,|A>|^{2}$. This can be refined if we have more knowledge. 
If it is known that one can go from $A$ to $C_{i}$ ($i=1,\cdots,n$)
and from $C_{i}$ to $B$ and that the intermediate states $|C_{i}>$ are a complete set of orthonormal alternatives then we can assume that 
$<C_{i}\,|C_{i}> = 1$ for each $i$ and that $\Sigma_{i} |C_{i}><C_{i}| = 1.$  This identity now corresponds to the fact that
$1$ is the sum of the probabilities of an arbitrary state being projected into one of these intermediate states.
\bigbreak

If there are intermediate states between the intermediate states this formulation can be continued
until one is summing over all possible paths from $A$ to $B.$ This becomes the path integral expression 
for the amplitude $<B|A>.$
\bigbreak

\subsection{Quantum Formalism and DNA Replication}
We wish to draw attention to the remarkable fact that this formulation of the expansion of 
intermediate quantum states has exactly the same pattern as our formal summary of DNA replication.
Compare them. The form of DNA replication is shown below. Here the environment of possible base pairs is represented by the ket-bra
$E = |C><W \,|.$

$$<W|C>\, \longrightarrow \, <W|\, |C>\, \longrightarrow \,<W| E |C>$$ 

$$\longrightarrow\, <W|\, |C><W|\, |C>\, \longrightarrow\, <W|C><W|C>$$

\noindent Here is the form of intermediate state expansion.

$$<B\,|A>\, \longrightarrow\, <B\,|\,|A>\, \longrightarrow \, <B\,|\,1 \,|A>$$

$$ \longrightarrow \, <B\,|\,\, \Sigma_{k}\,|C_{k}><C_{k}\,|\,\, |A> \, \longrightarrow \,\Sigma_{k}<B\,|C_{k}><C_{k}\,|A>.$$

\noindent We compare $$E = |C><W\,|$$ \noindent and $$1 = \Sigma_{k}\,|C_{k}><C_{k}\,|.$$

\noindent That the unit $1$ can be written as a sum over the intermediate states is an expression of how the 
environment (in the sense of the space of possibilities) impinges on the quantum amplitude, just as the expression of the environment as a 
soup of bases ready to be paired (a classical space of possibilities) serves as a description of the biological environment.
The symbol $E = |C><W\,|$ indicated the availability of the bases from the environment to form the 
complementary pairs. The projection operators $|C_{i}><C_{i}\,|$ are the possibilities for interlock of 
initial and final state through an intermediate possibility. In the quantum mechanics the special pairing is not of bases but
of a state and a possible intermediate from a basis of states. It is through this common theme of 
pairing that the conceptual notation of the bras and kets lets us see a correspondence between such
separate domains.
\bigbreak

\subsection{Quantum Copies are not Possible}
Finally, we note that in quantum mechanics it is not possible to copy a quantum state!
This is called the no-cloning theorem of elementary quantum mechanics \cite{Lom}. Here is the proof: 
\smallbreak

\noindent{\bf Proof of the No Cloning Theorem.} In order to have a quantum process make a copy of a quantum
state we need a unitary mapping $U:H \otimes H \longrightarrow H \otimes H$ where $H$ is a complex vector space such that there is a fixed state
$|X>\, \in H$ with the property that $$U(|X>|A>) = |A>|A>$$ \noindent for any state $|A> \in H.$ ($|A>|B>$ denotes
the tensor product $|A> \otimes |B>.$) Let
$$T(|A>) = U(|X>|A>) = |A>|A>.$$ \noindent Note that $T$ is a linear function of $|A>.$  Thus we have 

$$T|0> = |0>|0> = |00>,$$

$$T|1> = |1>|1>=|11>,$$

$$ T(\alpha |0> + \beta |1>) = (\alpha |0> + \beta |1>)(\alpha |0> + \beta |1>).$$ 

\noindent But

$$ T(\alpha |0> + \beta |1>) = \alpha |00> + \beta |11>.$$  \noindent Hence

$$\alpha |00> + \beta |11> =(\alpha |0> + \beta |1>)(\alpha |0> + \beta |1>)$$

$$ = \alpha^{2} |00> + \beta^{2} |11> + \alpha \beta |01> + \beta \alpha |10>$$

\noindent From this it follows that 
$\alpha \beta = 0.$ Since $\alpha$ and $\beta$ are arbitrary complex numbers, this is a contradiction. $\hfill \Box $
\bigbreak

The proof of the no-cloning theorem depends crucially on the linear superposition of quantum states and the linearity of quantum process.
By the time we reach the molecular level and attain the possibility of copying DNA molecules we are copying in a quite different sense than the 
ideal quantum copy that does not exist. The DNA and its copy are each quantum states, but they are different quantum states! That we see the two 
DNA molecules as identical is a function of how we filter our observations of complex and entangled quantum states. Nevertheless, the identity
of two DNA copies is certainly at a deeper level than the identity of the two letters ``i" in the word identity. The latter is conventional and symbolic.
The former is a matter of physics and biochemistry. 
\bigbreak

\section{Mathematical Structure and Topology}

We now comment on the conceptual underpinning for the notations and logical constructions that we 
use in this paper.  This line of thought will lead to topology and to the formalism for replication discussed 
in the last section.  
\bigbreak

Mathematics is built through distinctions, definitions, acts of language that bring forth logical 
worlds, arenas in which actions and patterns can take place. As far as we can determine at the present
time, mathematics while capable of describing the quantum world, is in its very nature quite 
classical. Or perhaps we make it so. As far as mathematics is concerned, there is no ambiguity in the
$1 + 1$ hidden in $2.$ The mathematical box shows exactly what is potential to it when it is opened.
There is nothing in the box except what is a consequence of its construction.
With this in mind, let us look at some mathematical beginnings.
\bigbreak

Take the beginning of set theory. We start with the empty set  $\phi = \{\,\,\,\}$ and we build new sets by the 
operation of set formation that takes any collection and puts brackets around it:

$$a\, b\, c\, d\, \longrightarrow  \{a,b,c,d\}$$

\noindent making a single entity $\{a,b,c,d \}$ from the multiplicity of the ``parts" that are so collected.
The empty set herself is the result of ``collecting nothing". The empty set is identical to the act of 
collecting. At this point of emergence the empty set is an action not a thing. Each subsequent
set can be seen as an action of collection, a bringing forth of unity from multiplicity. 
\bigbreak

One declares two sets to be the same if they have the same members. With this prestidigitation of language, the 
empty set becomes unique and a hierarchy of distinct sets arises as if from nothing.

$$\,\,\, \longrightarrow \{\,\,\,\} \longrightarrow \{\, \{\,\,\}\, \} \longrightarrow \{\, \{\,\,\} \, , \{\, \{\,\,\}\, \}\, \}
\longrightarrow \cdots$$

\noindent All representatives of the different mathematical cardinalities arise out of the void in the presence of these
conventions for collection and identification.
\bigbreak 

We would like to get underneath the formal surface. We would like to see what makes this formal hierarchy tick.
Will there be an analogy to biology below this play of symbols? On the one hand it is clear to us that there is
actually no way to go below a given mathematical construction. Anything that we call more fundamental will be 
another mathematical construct. Nevertheless, the exercise is useful, for it asks us to look closely at how this given 
formality is made.  It asks us to take seriously the parts that are usually taken for granted.
\bigbreak

We take for granted that the particular form of container used to represent the empty set 
is irrelevant to the empty set itself. But how can this be? In order to have a concept of emptiness, one 
needs to hold the contrast of that which is empty with ``everything else". One may object that these images are not
part of the formal content of set theory. But they are part of the {\em formalism} of set theory. 
\bigbreak

Consider the 
representation of the empty set: $\{\,\,\,\,\}.$  That representation consists in a bracketing that we take to indicate an
empty space within the brackets, and  an injunction to ignore the complex typographical domains outside the brackets.
Focus on the brackets themselves.  They come in two varieties: the left bracket,  $\{,$ and the right bracket, $\}.$  
The left bracket indicates a distinction of left
and right with the emphasis on the right. The right bracket indicates a distinction between left and right with an
emphasis on the left.  A left and right bracket taken together become a 
{\em container} when each is in the domain indicated by the other.  Thus in the bracket symbol 
$$\{ \, \, \, \}$$
\noindent for the empty set, the left bracket, being to the left of the right bracket, is in the left domain that is 
marked by the right bracket, and the right bracket, being to the right of the left bracket is in the right domain that 
is marked by the left bracket. The doubly marked domain between them is their content space, the arena of the empty set.
\bigbreak

The delimiters of the container are each themselves iconic for the process of making a distinction. In the notation of 
curly brackets,$\,\{\,$, this is particularly evident. The geometrical form of the curly bracket is a cusp singularity, the
simplest form of bifurcation. The relationship of the left and right brackets is that of a form and its mirror image.
If there is a given distinction such as left versus right, then the mirror image of that distinction is the one with 
the opposite emphasis. This is precisely the relationship between the left and right brackets. A form and its
mirror image conjoin to make a container.
\bigbreak

The delimiters of the empty set could be written in the opposite order:  $\}\{.$
This is an {\em extainer}. The extainer indicates regions external to itself. 
In this case of symbols on a line, the extainer
$\}\{$ indicates the entire line to the left and to the right of itself. The extainer is as natural as the container, but
does not appear formally in set theory. To our knowledge, its first appearance is in the Dirac notation of ``bras" and 
``kets" where Dirac takes an inner product written in the form $<B|A>$ and breaks it up into $<B\,|$ and $|A>$ and then
makes projection operators by recombining in the opposite order as $|A><B\,|.$  See the earlier discussion of quantum 
mechanics in this paper.
\bigbreak

Each left or right bracket in itself makes a distinction. The two brackets
are distinct from one another by mirror imaging,  which we take to be a
notational reflection of a fundamental process (of distinction) whereby
two forms are identical (indistinguishable) except by comparison in the
space of an observer. The observer {\em is} the distinction between the
mirror images.
Mirrored pairs of individual brackets interact to form either a {\em container}
$$C = \{\}$$ \noindent or an {\em extainer} $$E = \}\{.$$ 

\noindent These new forms combine to make:

$$CC = \{\}\{\} = \{E\}$$ \noindent and 

$$EE = \}\{\}\{=\}C\{.$$

\noindent Two containers interact to form an extainer within container brackets. Two extainers interact to form a container between
extainer brackets. The pattern of extainer interactions can be regarded as a formal generalization of the bra and ket patterns of the 
Dirac notation that we have used in this paper both for DNA replication and for a discussion of quantum mechanics. In the quantum mechanics
application $\{\}$ corresponds to the inner product $<A\,|B>$, a commuting scalar, while $\}\{$ corresponds to $|A><B\,|$, a matrix that 
does not necessarily commute with vectors or other matrices. With this application in mind, it is natural to decide to make the 
container an analog of a scalar quantity and let it commute with individual brackets. We then have the equation

$$EE = \}\{\}\{= \}C\{ = C \}\{ = CE.$$

\noindent By definition there will be no corresponding equation for $CC$.
We adopt the axiom that containers commute with other elements in this combinatorial algebra.
Containers and extainers are distinguished by this property. Containers appear as autonomous entities and can be moved about.
Extainers are open to interaction from the outside and are sensitive to their surroundings. 
At this point, we have described the basis for the formalism used in the earlier parts of this paper.
\bigbreak

If we interpret E as the ``environment" then the equation $\}\{ = E = 1$ expresses the availability of 
complementary forms so that 

$$\{ \} \longrightarrow \{ E \} \longrightarrow \{ \}\{ \}$$ 

\noindent becomes the form of DNA reproduction. 
\bigbreak

We can also regard $EE = \{\} E$ as symbolic  of the emergence of DNA from the chemical substrate. Just as the formalism
for reproduction ignores the topology, this formalism for emergence ignores the formation of the DNA backbone along 
which are strung the complementary base pairs. In the biological domain we are aware of levels of ignored structure.
\bigbreak

In mathematics it is customary to stop the examination of certain issues in order to 
create domains with requisite degrees of clarity. We are all aware that the operation of collection is proscribed 
beyond a certain point. For example, in set theory the Russell class $R$ of all sets that are not members of
themselves is not itself a set. It then follows that $\{R\},$ the collection whose member is the Russell class, is not a
class (since a member of a class is a set). This means that the construct $\{R\}$ is outside of the discourse of 
standard set theory. This is the limitation of expression at the ``high end" of the formalism. That the set theory has
no language for discussing the structure of its own notation is the limitation of the language at the ``low end".
Mathematical users, in speaking and analyzing the mathematical structure, and as its designers, can speak beyond both
the high and low ends.
\bigbreak

In biology we perceive the pattern of a formal system, a system that is embedded in a structure whose complexity 
demands the elucidation of just those aspects of symbols and signs that are commonly ignored in the mathematical 
context. Rightly these issues should be followed to their limits. The curious thing is what peeks through when we just
allow a bit of it, then return to normal mathematical discourse. With this in mind, lets look more closely at the 
algebra of containers and extainers.
\bigbreak

Taking two basic forms of bracketing, an intricate algebra appears from their elementary
interactions:

$$E =\, ><$$
$$F =\, ][$$
$$G =\, >[$$
$$H =\, ]<$$

\noindent are the extainers, with corresponding containers:

$$<>, \,\,\,\,\, [], \,\,\,\,\, [>, \,\,\,\,\, <].$$

\noindent These form a closed algebraic system with the following multiplications:

$$EE =\, ><\,>< =\, <> E$$
$$FF =\, ][\,][ =\, [] F$$
$$GG =\, >[\,>[ =\, [> G$$
$$HH =\, ]<\,]< =\, <] H$$

\noindent and

$$EF =\, ><\,][ =\, <] G$$
$$EG =\, ><\,>[ =\, <> G$$
$$EH =\, ><\,]< =\, <] E$$

$$FE =\, ][\,>< =\, [> H$$
$$FG =\, ][\,>[ =\, [> F$$
$$FH =\, ][\,]< =\, [] H$$

$$GE =\, >[\,>< =\, [> E$$
$$GF =\, >[\,][ =\, [] G$$
$$GH =\, >[\,]< =\, [] E$$

$$HE =\, ]<\,>< =\, <> H$$
$$HF =\, ]<\,][ =\, <] F$$
$$HG =\, ]<\,>[ =\, <> F$$

\noindent Other identities follow from these. For example,

$$EFE =\, ><][>< =\, <][> E.$$

This algebra of extainers and containers is a precursor to the Temperley Lieb algebra, an algebraic structure that 
first appeared (in quite a different way) in the study of the Potts model in statistical mechanics \cite{Baxter}.
We shall forgo here details about the Temperley Lieb algebra itself, and refer the reader to \cite{QCJP} where this point 
of view is used to create unitary representations of that algebra for the context of quantum computation.
Here we see the elemental nature of this algebra, and how it comes about quite naturally once one adopts a formalism
that keeps track of the structure of boundaries that underlie the mathematics of set theory.
\bigbreak

The {\em Temperley Lieb algebra} $TL_{n}$ is an algebra over a commutative ring $k$ with
generators $\{ 1, U_{1},U_{2}, ... ,U_{n-1} \}$ and relations 

$$U_{i}^{2} = \delta U_{i},$$

$$U_{i}U_{i \pm 1}U_{i} = U_{i},$$

$$U_{i}U_{j} = U_{j}U_{i}, |i-j|>1,$$

\noindent where $\delta$ is a chosen element of the ring $k$. These equations give the
multiplicative structure of the algebra. The algebra is a free module over the ring $k$ with basis
the equivalence classes of these products modulo the given relations.
\bigbreak

To match this pattern with our combinatorial algebra let $n=2$ and let $U_{1} = E = ><$, $U_{2} = F = ][$ and assume that 
$1 = <] = [>$ while $\delta = <> = [].$ The above equations for our combinatorial algebra then match the multiplicative equations of the
Temperley Lieb algebra. 
\bigbreak

The next stage for representing the Temperley Lieb algebra is a diagrammatic representation that uses two different forms of 
extainer. The two forms are obtained not by changing the shape of the given extainer, but rather by shifting it 
relative to a baseline. Thus we define diagrammatically $U = U_{1}$ and $V = U_{2}$ as shown below:

$$U =\begin{array}{c}
 --\\
 > <
\end{array}$$

$$V = \begin{array}{c}
> <\\
--
\end{array} 
$$

$$UU = \begin{array}{c}
 ----\\
 ><><
\end{array} 
= <> \begin{array}{c}
 --\\
 ><
\end{array} 
= <> U$$

$$UVU = 
\begin{array}{c}
     _{^{---}>\,\,<^{---}}\\
     ^{>\,\,\,<_{--}>\,\,\,<}
\end{array} 
= \begin{array}{c}
 _{^{----}}  \\ 
^{>\,\,\,\,\,<}
\end{array} 
= U.$$

\noindent In this last equation $UVU = U$ we have used the topological deformation of the connecting line from top to top to obtain
the identity. In its typographical form the identity requires one to connect corresponding endpoints of the brackets. In Figure 2
we indicate a smooth picture of the connection situation and the corresponding topological deformation of the lines. 
We have deliberately shown the derivation in a typographical mode to emphasize its essential difference from the 
matching pattern that produced $$EFE =\, ><][>< =\, <][> E.$$ By taking the containers and extainers shifted this way, we enter a new and
basically topological realm. This elemental relationship with topology is part of a deeper connection where the Temperley Lieb algebra is used to 
construct representations of the Artin Braid Group.  This in turn leads to the construction of the well-known Jones
polynomial invariant of knots and links via the bracket state model \cite{KS}. It is not the purpose of this paper to go into the details of those 
connections, but rather to point to that place in the mathematics where basic structures apply to biology, topology, 
and logical foundations. 
\bigbreak

$$ \picill4inby4in(Figure2)  $$

\begin{center}
{\bf Figure 2 - A Topological Identity }
\end{center}

It is worthwhile to point out that the formula for expanding the bracket polynomial can be indicated symbolically in the same fashion that
we used to create the Temperley Lieb algebra via containers and extainers.
We will denote a crossing in the link diagram by the
letter chi, \mbox{\large $\chi$}. The
letter itself denotes a crossing where {\em the curved line in the letter chi is crossing over the straight segment in the
letter}. The barred letter denotes the switch of this crossing where {\em the curved line in the letter chi is undercrossing
the straight segment in the letter}. In the bracket state model a crossing in a diagram for the
knot or link is expanded into two possible states by either smoothing (reconnecting) the crossing horizontally, \mbox{\large
$\asymp$}, or vertically $><$.  The vertical smoothing can be regarded as the extainer and the horizontal smoothing as an identity operator.
In a larger sense, we can regard both smoothings as extainers with different relationships to their environments. In this sense the crossing 
is regarded as the superposition of horizontal and vertical extainers.
The crossings expand
according to the formulas  
$$\mbox{\large $\chi$} = A \mbox{\large $\asymp$} + A^{-1} ><$$
$$\overline{\mbox{\large $\chi$}} = A^{-1} \mbox{\large $\asymp$} + A ><.$$
\noindent The verification that the bracket is invariant under the second Reidemeister move is then seen by verifying that
$$\mbox{\large $\chi$}\overline{\mbox{\large $\chi$}} = \mbox{\large $\asymp$}.$$ For this one needs that the container $<>$ has value
$-A^2 - A^{-2}$ (the loop value in the model). The significant mathematical move in producing this model is the notion of the crossing as a 
superposition of its smoothings.
\bigbreak

It is useful to use the iconic symbol $><$ for the extainer, and to choose another iconic symbol  \mbox{\large $\asymp$}
for the identity operator in the algebra. With these choices we have
$$\mbox{\large $\asymp$}\mbox{\large $\asymp$} \,\, = \,\, \mbox{\large $\asymp$}$$
$$\mbox{\large $\asymp$} >< \,\, = \,\, >< \mbox{\large $\asymp$} \,\,= \,\, ><$$
\bigbreak

\noindent Thus
$$\mbox{\large $\chi$}\overline{\mbox{\large $\chi$}}$$
$$ =( A \mbox{\large $\asymp$} + A^{-1} ><)(A^{-1} \mbox{\large $\asymp$} + A ><)$$ 
$$= AA^{-1} \mbox{\large $\asymp$}\mbox{\large $\asymp$} + A^{2}\mbox{\large $\asymp$}>< + A^{-2}><  \mbox{\large $\asymp$}
+ AA^{-1} ><><$$
$$=  \mbox{\large $\asymp$}  + A^{2}>< + A^{-2}>< +  \delta ><$$
$$ = \mbox{\large $\asymp$}  + (A^{2} + A^{-2} +  \delta) >< $$
$$=  \mbox{\large $\asymp$}$$

\noindent Note the use of the extainer identity $><>< = >\, \delta\, < = \delta \, ><.$
At this stage the combinatorial algebra of containers and extainers emerges as the background to the topological characteristics of the 
Jones polynomial.
\bigbreak

\subsection{Protein Folding and Combinatorial Algebra}
The approach in this section derives from ideas in \cite{KM}.
Here is another use for the formalism of bras and kets. Consider a molecule that is obtained by ``folding" a long chain molecule.
There is a set of sites on the long chain that are paired to one another to form the folded molecule. The difficult problem in protein 
folding is the determination of the exact form of the folding given a multiplicity of possible paired sites. Here we assume that the 
pairings are given beforehand, and consider the abstract structure of the folding {\em and} its possible embeddings in three dimensional
space. {\em Let the paired sites on the long chain be designated by labeled bras and kets with the bra appearing before the ket in the 
chain order.}  Thus $<A|$  and $|A>$ would denote such a pair and 
the sequence  $$C = <a|<b|<c||c>|b><d||d>|a><e||e>$$ \noindent could denote the paired sites on the long chain. See Figure 3 for 
a depiction of this chain and its folding.
In this formalism we do not assume any identities about moving containers or extainers, since the 
exact order of the sites along the chain is of great importance. We say that two chains are {\em isomorphic} if they differ only in
their choice of letters. Thus $<a|<b||b>|a>$ and $<r|<s||s>|r>$ are isomorphic chains. Note that each bra ket pair in a chain is decorated
with a distinct letter.
\bigbreak

Written in bras and kets a chain has an underlying parenthesis structure
that is obtained by removing all vertical bars and all letters. Call this $P(C)$ for a given chain $C$. Thus we have

$$P(C) = P(<a|<b|<c||c>|b><d||d>|a><e||e>) = <<<>><>><>.$$

Note that in this case we have $P(Chain)$ is a legal parenthesis structure in the usual sense of containment and paired brackets.
Legality of parentheses is defined inductively:
\begin{enumerate}
\item $<>$ is legal.
\item If $X$ and $Y$ are legal, then $XY$ is legal.
\item If $X$ is legal, then $<X>$ is legal.
\end{enumerate}
\noindent These rules define legality of finite parenthetic expresssions.
In any legal parenthesis structure, one can deduce directly from that structure which brackets are paired with one another. 
Simple algorithms suffice for this, but we omit the details. In any case a legal parenthesis structure has an intrinsic pairing associated with it,
and hence there is an inverse to the mapping $P$. We define $Q(X)$ for $X$ a legal parenthesis structure, to be the result of
replacing each pair $\cdots < \cdots > \cdots$ in X by $\cdots <A| \cdots |A> \cdots$ where $A$ denotes a specific letter chosen for that 
pair, with different pairs receiving different letters. Thus $Q(<<>>) = <a|<b||b>|a>.$ Note that in the case above, we have that 
$Q(P(C))$ is isomorphic to $C.$
\bigbreak

$$ \picill4inby4in(Figure3)  $$

\begin{center}
{\bf Figure 3 - Secondary Structure $<a|<b|<c||c>|b><d||d>|a><e||e>$}
\end{center}

A chain $C$ is said to be a {\em secondary folding structure} if $P(C)$ is legal and $Q(P(C))$ is isomorphic to $C.$  
The reader may enjoy the exercise of seeing that 
secondary foldings (when folded) form tree-like structures without any loops or knots. This notion of secondary folding structure
corresponds to the usage in molecular biology, and it is a nice application of the bra ket formalism. This also shows the very rich
combinatorial background in the bras and kets that occurs before the imposition of any combinatorial algebra.
\bigbreak

Here is the simplest non-secondary folding: $$L = <a|<b||a>|b>.$$
\noindent Note that $P(L) =<<>>$ is legal, but that $Q(P(L)) = Q(<<>>) =<a|<b||b>|a>$ is not isomorphic to $L.$  
$L$ is sometimes called a ``pseudo knot" in the literature of protein folding. Figure 4 should make clear this nomenclature.
The molecule is folded back on itself in a way that looks a bit knotted.
\bigbreak

$$ \picill4inby4in(Figure4)  $$

\begin{center}
{\bf Figure 4 - A Tertiary Structure - $<a|<b||a>|b>$}
\end{center}

With these conventions it is convenient to abbreviate a chain by just giving its letter sequence and removing the (reconstructible) bras and kets.
Thus $C$ above may be abbreviated by $abccbddaee.$
\bigbreak

One may wonder whether at least theoretically there are foldings that would necessarily be knotted when embedded in three dimensional space.
With open ends, this means that the structure folds into a graph such that there is a knotted arc in the graph for some traverse from one end to 
the other. Such a traverse can go along the chain or skip across the bonds joining the paired sites.
The answer to this question is yes, there are folding patterns that can force knottedness. Here is an example of such an 
intrinsically knotted folding.

$$ABCDEFAGHIJKBGLMNOCHLPQRDIMPSTEJNQSUFKORTU.$$

It is easy to see that this string is not a secondary structure. To see that it is intrinsically knotted, we appeal to the Conway-Gordon Theorem
\cite{CG} that tells us that the complete graph on seven vertices is intrinsically knotted. In closed circular form (tie the ends of
the folded string together), the folding that 
corresponds to the above string retracts to the complete graph on seven vertices. Consequently, that folding, however it is embedded, must contain a knot
by the Conway-Gordon Theorem. We leave it as an exercise for the reader to draw an embedding corresponding to a folding of 
this string and to locate the knot!
The question of intrinsically knotted foldings that occur in nature remains to be investigated.
\bigbreak

\section{Cellular Automata}
Some examples from cellular automata clarify many of the issues about replication and the relationship of logic and biology. 
Here is an example due to 
Maturana, Uribe and Varela \cite{MUV}. See also \cite{FV} for a global treatment of related issues.
The ambient space is two dimensional and in it there are
``molecules" consisting in ``dots" (See Figure 5). There is a minimum distance between the 
dots (one can place them on a discrete lattice in the plane). And ``bonds" can form with
a probability of creation and a probability of decay between molecules with minimal
spacing. There are two types of molecules: ``substrate" and ``catalysts". The catalysts are not
susceptible to bonding, but their presence (within say three minimal step lengths) enhances the 
probability of bonding and decreases the probability of decay. Molecules that are not bonded
move about the lattice (one lattice link at a time) with a probability of motion.
In the beginning there is a randomly placed soup of molecules with a high percentage of substrate 
and a smaller percentage of catalysts. What will happen in the course of time?
\bigbreak

$$ \picill4inby3in(Figure5)  $$

\begin{center}
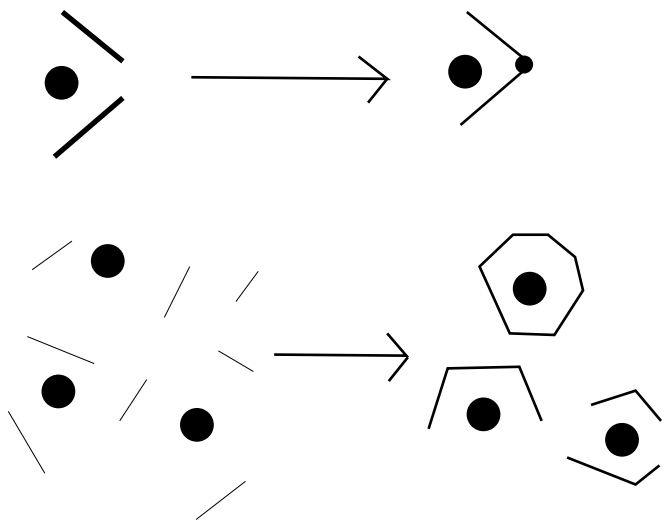
{\bf Figure 5  - Proto-Cells of Maturana, Uribe and Varela }
\end{center}

In the course of time the catalysts (basically separate from one another due to lack of bonding) 
become surrounded by circular forms of bonded or partially bonded substrate. A distinction 
(in the eyes of the observer) between inside (near the catalyst) and outside 
(far from a given catalyst) has spontaneously arisen through the ``chemical rules". Each catalyst
has become surrounded by a proto-cell. No higher organism has formed here, but there is a hint of the 
possibility of higher levels of organization arising from a simple set of rules of interaction.
{\em The system is not programmed to make the proto-cells.} They arise spontaneously in the evolution of 
the structure over time.
\bigbreak

One might imagine that in this way, organisms could be induced to arise as the evolutionary 
behavior of formal systems. There are difficulties, not the least of which is that there are 
nearly always structures in such systems 
whose probability of spontaneous emergence is vanishingly small. A good example is given by another 
automaton --  John H. Conway's ``Game of Life".  In ``Life" the cells  appear and disappear as
marked squares in a rectangular planar grid. A newly marked cell is said to be ``born". An unmarked cell
is ``dead". A cell dies when it goes from the marked to the unmarked state. A marked cell 
survives if it does not become unmarked in a given time step.
According to the rules of Life, an unmarked cell is born if and only if it has three neighbors.
A marked cell survives if it has either two or three neighbors. All cells in the lattice are updated in 
a single time step. The Life automaton is one of many automata of this type and indeed it is 
a fascinating exercise to vary the rules and watch a panoply of different behaviors. For this 
discussion we concentrate on some particular features. There is a configuration in Life called a
``glider". See Figure 6. This illustrates a ``glider gun" (discussed below) that produces a series of gliders
going diagonally from left to right down the Life lattice.
 The glider consists in five cells in one of two basic configurations.
Each of these configurations produces the other (with a change in orientation). After four steps the
glider reproduces itself in form, but shifted in space. Gliders appear as moving entities in 
the temporality of the Life board. The glider is a complex entity that arises naturally from a 
small random selection of marked cells on the Life board. Thus the glider is a ``naturally 
occurring entity" just like the proto-cell in the Maturana-Uribe-Varela automaton. But Life 
contains potentially much more complex phenomena. For example, there is the ``glider gun" (See
Figure 6) which perpetually creates new gliders.  The ``gun" was invented by a group of researchers
at MIT in the 1970's (The Gosper Group). It is highly unlikely that a gun would appear 
spontaneously in the Life board. Of course there is a tiny probability of this, but we would guess
that the chances of the appearance of the glider gun by random selection or evolution from a 
random state is similar to the probability of all the air in the room collecting in one corner.
Nervertheless, the gun is a natural design based on forms and patterns that do appear spontaneously 
on small Life boards.  The glider gun emerged through the coupling of the power of human cognition and the automatic behavior of 
a mechanized formal system.  
Cognition is in fact an attribute of our biological system at an 
appropriately high level of organization. But cognition itself looks as improbable as the glider gun!  
Do patterns as complex as cognition or the glider gun arise 
spontaneously in an appropriate biological context? 
\bigbreak

$$ \picill5.5inby4.5in(Figure6)  $$

\begin{center}
{\bf Figure 6  - Glider Gun and Gliders }
\end{center}

There is a middle ground.  If one examines cellular automata of a given type and varies the
rule set randomly rather than varying the initial conditions for a given automaton, then a very wide
variety of phenomena will present themselves. In the case of molecular biology at the level of the 
DNA there is exactly this possibility of varying the rules in the sense of varying the sequences in 
the genetic code. So it is possible at this level to produce a wide range of remarkable complex systems.
\bigbreak

\subsection{Other Forms of Replication}
Other forms of self-replication are quite revealing. For example, one might point out that a
stick can be made to reproduce by breaking it into two pieces. This may seem satisfactory on the 
first break, but the breaking cannot be continued indefinitely. In mathematics on the other hand, 
we can divide an interval into two intervals and continue this process ad infinitum. For a 
self-replication to have meaning in the physical or biological realm there must be a genuine 
repetition of structure from original to copy. At the very least the interval should grow to 
twice its size before it divides (or the parts should have the capacity to grow independently).
\bigbreak

A clever automaton, due to Chris Langton, takes the initial form of a square in the plane.
The rectangle extrudes a edge that grows to one edge length and a little more, 
turns by ninety degrees, grows one edge length, turns by ninety degrees grows one edge length,
turns by ninety degrees and when it grows enough to collide with the original extruded edge, 
cuts itself off to form a new adjacent square, thereby reproducing itself. This scenario is then 
repeated as often as possible producing a growing cellular lattice. See Figure 7. 
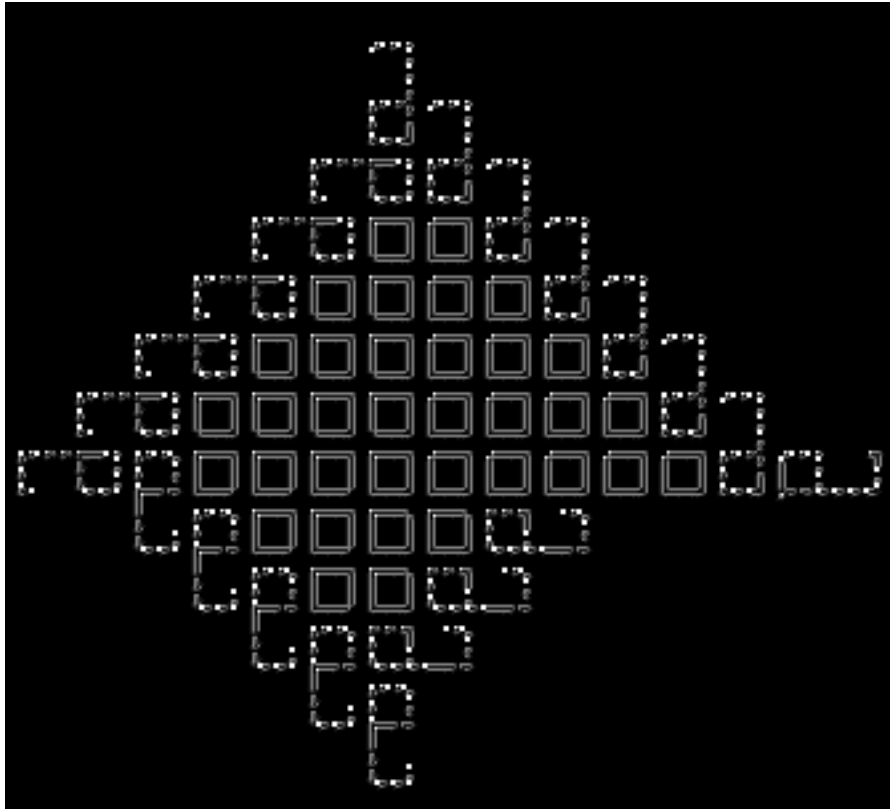
\bigbreak

$$ \picill5.5inby4.5in(Figure7)  $$

\begin{center}
{\bf Figure 7 - Langton's Automaton }
\end{center}

The replications that happen in automata such as Conway's Life are all really instances of periodicity
of a function under iteration. The gilder is an example where the Life game function $L$ applied to an
initial condition $G$ yields $L^{5}(G) = t(G)$ where $t$ is a rigid motion of the plane. Other intriguing
examples of this phenomenon occur. For example the initial condition $D$ for Life shown in Figure 8 has the property that 
$L^{48}(D) = s(D) + B$ where $s$ is a rigid motion of the plane and $s(D)$ and the residue $B$ are disjoint
sets of marked squares in the lattice of the game. $D$ itself is a small configuration of eight marked
squares fitting into a rectangle of size $4$ by $6.$ Thus $D$ has a probability of $1/735471$ of being chosen
at random as eight points from $24$ points.  
\bigbreak

$$ \picill1inby1in(Figure8)  $$

\begin{center}
{\bf Figure 8 - Condition D with geometric period $48$ }
\end{center}

Should we regard self-replication as simply an instance of periodicity under iteration?
Perhaps, but the details are more interesting in a direct view.  The glider gun in Life is a
structure $GUN$ such that $L^{30}(GUN) = GUN + GLIDER.$ Further iterations move the disjoint glider away
from the gun so that it can continue to operate as an initial condition for $L$ in the same way.
A closer look shows that the glider is fundamentally composed of two parts $P$ and $Q$ such that 
$L^{10}(Q)$ is a version of $P$ and some residue and such that $L^{15}(P) = P^* + B$ where $B$ is a rectangular 
block, and $P^*$ is a mirror image of $P$, while $L^{15}(Q) = Q^* + B'$ where $B'$ is a small non-rectangular
residue. See Figure 9 for an illustration showing the parts $P$ and $Q$ (left and right) flanked by small blocks that
form the ends of the gun. One also finds that $L^{15}( B + Q^*) = GLIDER + Q + Residue.$ This is the internal 
mechanism by which the glider gun produces the glider. The extra blocks at either end of the glider
gun act to absorb the residues that are produced by the iterations. Thus the end blocks are catalysts
that promote the action of the gun. Schematically the glider production goes as follows:

$$P+Q \longrightarrow P^* + B + Q*$$

$$B + Q^* \longrightarrow GLIDER + Q$$

\noindent whence

$$P+Q \longrightarrow P^* + B + Q^* \longrightarrow P + GLIDER + Q = P + Q + GLIDER.$$

\noindent The last equality symbolizes the fact that the glider is an autonomous entity no longer involved 
in the structure of $P$ and $Q.$ It is interesting that $Q$ is a spatially and time shifted version of 
$P.$ Thus $P$ and $Q$ are really ``copies" of each other in an analogy to the structural relationship of 
the Watson and Crick strands of the DNA. The remaining part of the analogy is the way the catalytic
rectangles at the ends of the glider gun act to keep the residue productions from interfering with 
the production process. This is analogous to the enzyme action of the topoisomerase in the DNA.
\bigbreak

$$ \picill4inby1.5in(Figure9)  $$

\begin{center}
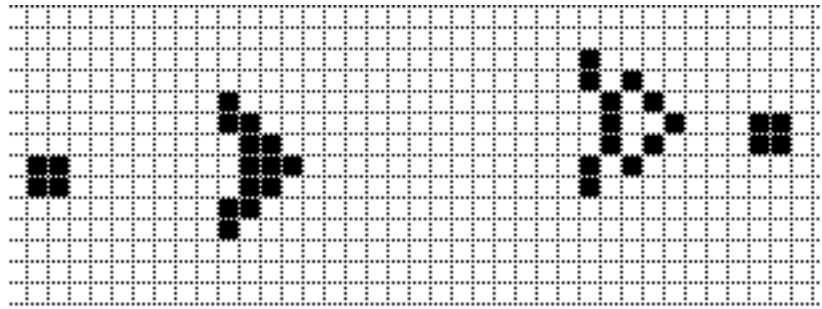
{\bf Figure 9 - P(left) and Q(right) Compose the Glider Gun }
\end{center}

The point about this symbolic or symbiological analysis is that it enables us to take an analytical 
look at the structure of different replication scenarios for comparison and for insight.
\bigbreak

\section{Epilogue - Logic and Biology}

We began with the general question: What is the relationship of logic and biology.
Certain fundamentals, common to both are handled quite differently. These  are certain 
fundamental distinctions: The distinction of symbol and object (the name and the thing that is named).
The distinction of a form and a copy of that form.  
\bigbreak

In logic the symbol and its referent are normally taken to be distinct. This leads to a host of 
related distinctions such as the distinction between a description or blueprint and the object 
described by that blueprint.  A related distinction is the dichotomy between software and hardware.
The software is analogous to a description. Hardware can be constructed with the aid of a blueprint or
description. But software intermediates between these domains as it is an {\em instruction.} An instruction is not
a description of a thing, but a blueprint for a process. Software needs hardware in order to 
become an actual process. Hardware needs software as a directive force. Although mutually dependent,
hardware and software are quite distinct.
\bigbreak

In logic and computer science the boundary between hardware and software is first met at the machine 
level with the built-in capabilities of the hardware determining the type of software that can be 
written for it. Even at the level of an individual gate, there is the contrast of the 
structure of that gate as a design and the implementation of that design that is used in the
construction of the gate. The structure of the gate is mathematical. Yet there is the physical implementation of these
designs, a realm where the decomposition into parts is not easily 
mutable. Natural substances are used, wood, metal, particular compounds, atomic elements and so on.
These are subject to chemical or even nuclear analysis and production, but eventually one 
reaches a place where Nature takes over the task of design.
\bigbreak

In biology it is the reverse. No human hand has created these designs. The organism stands for itself,
and even at the molecular level the codons of the DNA are not symbols. They do not stand for 
something other than themselves. They cooperate in a process of production, but no one wrote their sequence as
software. There is no software. There is no distinction between hardware and software in 
biology. 
\bigbreak

In logic a form arises via the syntax and alphabet of a given formal system.  
That formal system arises via the choices of the mathematicians who create it. They create it through
appropriate abstractions. Human understanding fuels the operation of a formal system.
Understanding imaged into programming fuels the machine operation of a mechanical image of that formal 
system. The fact that both humans and machines can operate a given formal system has lead to much confusion, for they
operate it quite differently. 
\smallbreak

\noindent {\em Humans are always on the edge of breaking the rules either through error or inspiration.
Machines are designed by humans to follow the rules, and are repaired when they do not do so. Humans are encouraged to operate 
through understanding, and to create new formal systems (in the best of all possible worlds).} 
\smallbreak

\noindent Here is the ancient 
polarity of syntax (for the machine) and semantics (for the person). The person must mix syntax and semantics to come to
understanding. So far, we have only demanded an adherence to syntax from the machines.
\bigbreak

The movement back and forth between syntax and semantics underlies all attempts to
create logical or mathematical form. This is the cognition behind a given formal 
system. There are those who would like to create cognition on the basis of syntax alone.
But the cognition that we all know is a byproduct or an accompaniment to biology. 
Biological cognition comes from a domain where there is at base no distinction between syntax 
and semantics. To say that there is no distinction between syntax and semantics in biology
is not to say that it is pure syntax. Syntax is born of the possibility of such a 
distinction.
\bigbreak

In biology an energetic chemical and quantum substrate gives rise to a ``syntax" of combinational 
forms (DNA, RNA, the proteins, the cell itself, the organization of cells into the organism).
These combinational forms give rise to cognition in human organisms. Cognition gives rise to
the distinction of syntax and semantics. Cognition gives rise to the possibility of design,
measurement, communication, language, physics and technology.
\bigbreak

In this paper we have covered a wide ground of ideas related to the foundations of mathematics and its relationship with biology and with 
physics. There is much more to explore in these domains. The result of our exploration has been the articulation of a mathematical region that 
lies in the crack between set theory and its notational foundations. We have articulated the concepts of container $<>$ and extainer $><$
and shown how the formal algebras generated by these forms encompass significant parts of the logic of DNA replication, the Dirac formalism for
quantum mechanics, formalism for protein folding and the Temperley Lieb algebra at the foundations of topological invariants
of knots and links.  It is the 
mathematician's duty to point out formal domains that apply to a multiplicity of contexts. In this case we suggest that it is just possible that 
there are deeper connections among these apparently diverse contexts that are only hinted at in the steps taken so far. The common formalism 
can act as compass and guide for further exploration.
\bigbreak

\bigbreak

\noindent {\sc L.H.Kauffman: Department of Mathematics, Statistics and Computer Science, University
of Illinois at Chicago, 851 South Morgan St., Chicago IL 60607-7045, U.S.A.}

\vspace{.1in}
\noindent {\sc E-mail: \ kauffman@math.uic.edu }


\begin{thebibliography}{131}


\bibitem{Baxter} R. J. Baxter, ``Exactly Solved Models in Statistical Mechanics", {\em Academic Press}, 1982.

\bibitem{BD} H.P. Barendregt, ``The Lambda Calculus Its Syntax and Semantics",{\em North Holland}, 1981 and 1985.

\bibitem{WB} W. Bricken and E. Gullichsen, An Introduction to Boundary Logic with the Losp Deductive Engine,
 {\em Future Computing Systems 2(4)},(1989) 1-77.

\bibitem{CG} J. H. Conway and C. McA Gordon, Knots and links in spatial graphs, {\em J. Graph Theory}, Vol. 7 (1983), 445-453.

\bibitem{D} P. A. M. Dirac, ``Principles of Quantum Mechanics", {\em Oxford University Press} 1958. 

\bibitem{KH} J. E. Hearst, L.H. Kauffman, and W. M. McClain, A simple mechanism
for the avoidance of entanglement during chromosome replication,  {\em Trends in
Genetics}, June 1998, Vol. 14, No. 6, 244-247.

\bibitem{KL} L. H. Kauffman, Knot Logic,  In ``Knots and Applications"  ed. by L. Kauffman, {\em World Scientific Pub. Co.},
(1994), 1-110.

\bibitem{KM} L. Kauffman and Y. Magarshak,  Graph invariants and the topology of RNA folding, {\em  Journal
of Knot Theory and its Ramifications}, Vol3, No.3, 233-246.

\bibitem{KS} L. H. Kauffman, State Models and the Jones Polynomial,  Topology 26 (1987), 395-407.

\bibitem{GSB} G. Spencer-Brown, ``Laws of Form", {\em Julian Press}, New York (1969).

\bibitem{LD} E. L. Zechiedrich, A. B. Khodursky, S. Bachellier, R. Schneider, D. Chen, D. M. J. Lilley and 
N. R. Cozzarelli, Roles of topoisomerases in maintaining steady-state DNA supercoiling in {\em Escherichia coli},
{\em J. Biol. Chem.} 275:8103-8113 (2000).

\bibitem{Lom} S. Lomonaco Jr, A Rosetta Stone for Quantum
Mechanics with an Introduction to Quantum Computation, in
``Quantum Computation: A Grand Mathematical Challenge for the
Twenty-First Century and the Millennium," AMS, Providence, RI (2000)
(ISBN 0-8218-2084-2).

\bibitem{MUV} H. R. Maturana, R. Uribe and F. G. Varela, Autopoesis: The organization of living systems, its characterization and 
a model, {\em Biosystems}, Vol. 5, (1974), 7-13.

\bibitem{QCJP} L. H. Kauffman, Quantum Computing and the Jones Polynomial (to appear in ``Proceedings of AMS special session on
Quantum Computing", edited by S. Lomonaco Jr).

\bibitem{FV}	F. J. Varela,  ``Foundations of Biological Autonomy", {\em North Holland Press} (1979).

\end{thebibliography}
\end{document}